\documentclass[twocolumn,fleqn,natbib]{svjour2}
\bibpunct{[}{]}{;}{n}{}{,} 

\smartqed  
\usepackage{graphicx}
\usepackage{amsmath}
\journalname{Granular Matter}
\begin{document}

\title{High intensity tapping regime in a frustrated lattice gas model of granular compaction.}

\titlerunning{High intensity tapping...}        

\author{Paula A. Gago \and Nicol\'{a}s E. Bueno \and Luis A. Pugnaloni
} \institute{Paula A. Gago: Departamento de F\'{\i}sica, Facultad de Ciencias Exactas, Universidad 
             Nacional de La Plata, Casilla de Correo 67, 1900 La Plata, Argentina.
               \\
             Nicol\'{a}s E. Bueno: Laboratorio de Ense\~{n}anza de la F\'{\i}sica, Departamento de
             F\'{\i}sica, Facultad de Ciencias Exactas, Universidad Nacional de La Plata, Casilla de
             Correo 67, 1900 La Plata, Argentina.
               \\
             L. A. Pugnaloni: Instituto de F\'{\i}sica de L\'{\i}quidos y Sistemas Biol\'{o}gicos
             (CONICET La Plata - UNLP), Casilla de Correo 565, 1900 La Plata, Argentina. Tel.:
             +54-221-4233283, Fax: +54-221-4257317, E-mail: luis@iflysib.unlp.edu.ar
}

\authorrunning{Paula A. Gago et al.} 

\date{Received: date }

\maketitle

\begin{abstract}
In the frame of a well established lattice gas model for granular compaction, we investigate the high intensity tapping regime where a pile expands significantly during external excitation. We find that this model shows the same general trends as more sophisticated models based on molecular dynamic type simulations. In particular, a minimum in packing fraction as a function of tapping strength is observed in the reversible branch of an annealed tapping protocol.

\keywords{Frustrated Lattice Gas \and Granular Compaction \and Tapping \and Reversible Branch}
\end{abstract}

\section{Introduction}
\label{intro}
The phenomenology of compaction of granular matter under vertical tapping captures the attention of a number of investigations and is not exempt from controversies \cite{Richard1}. Most studies nowadays focus on the dynamics of compaction---the evolution of structural properties as a function of the number of taps applied to the sample. In spite of being of chief importance, much less work is done on the steady state regime achieved by the sample after a very large number of taps. At low tapping intensities, the relaxation dynamics of these systems is extremely slow, which makes the steady state very hard to reach. In a pioneering work, Nowak et al. \cite{Nowak1} showed that the steady state can be achieved by means of a suitable annealing. Recently, Ribi\`{e}re et al. \cite{Ribiere1} argued that the steady state is indeed obtainable, reproducible and may constitute a true thermodynamic state for granular systems. These experiments show that the packing fraction $\phi$ in the steady state is a monotonic decreasing function of the tapping intensity: the so-called ``reversible branch''. A number of studies, from experimental \cite{Nowak1,Ribiere1,Schroter1} and modeling \cite{Stadler1,Philippe1,Cimarra1,Mehta1,Nicodemi1,Nicodemi2,Pugnaloni3} approaches have confirmed this general trend. One exception has been found in a model of pentagon packings \cite{Vidales1,Vidales2} where the reversible branch presents a monotonic increase of $\phi$ as tapping intensity is increased. 

Recently, by using a few models of granular deposition, it has been shown \cite{Pugnaloni1} that $\phi$, in the reversible branch, presents a minimum at relatively high tapping intensities. This has now been observed in the laboratory \cite{Sanchez1}. Previous experimental studies where unable to observe this feature presumably due to the high intensity taps required. According to the simulation models, it may be necessary to transfer enough energy to the packing during a tap to induce the system to expand up to five times its volume before this minimum can be observed. This is a rather difficult experiment to carry out with a setup designed to explore slow relaxation by gentle tapping. In twodimensional systems, the minimum can be observed at much lower tapping strength; an effective expansion during the tap of less than twice the bed height. It is worthmentioning that some of the models used in Ref.~\cite{Pugnaloni1} where the same models used by others \cite{Philippe1,Mehta1}, but focusing in a different region of the parameters that control tap intensity.

The use of annealed tapping to investigate the hysteresis of granular assemblies subjected to tapping was also used in the frame of a very simple frustrated lattice gas model proposed by Nicodemi et al. \cite{Nicodemi1,Nicodemi2} to investigate granular compaction. In this work we assess the ability of this model to display a reversible branch similar to the one observed in experiments and whether the density minimum predicted by more sophisticated models is also present. With this aim, we explore a range of parameter values of the model not yet reported in the literature in order to simulate strong intensity tapping conditions.

The rest of the paper is organized as follows. In Sect.~\ref{sec:1} we review the main features of the frustrated lattice gas model \cite{Nicodemi1}. In Sect.~\ref{sec:2} we describe the tapping protocol. In Sect.~\ref{sec:3} we present the results. In Sect.~\ref{sec:4} we draw the conclusions.

\section{Frustrated lattice gas model}
\label{sec:1}
We consider a model first proposed by Nicodemi et al. \cite{Nicodemi1}. The model consists of a
system of particles that move on a square lattice whose bonds are characterized by fixed random numbers $\epsilon_{i,j}=\pm 1$. Any site $i$ can be either empty or filled with a particle. Particles are characterized by an internal degree of freedom (spin) $S_i=\pm 1$ and are subjected to the constraint that
whenever two neighboring sites $i$ and $j$ hold particles with spin $S_i$ and $S_j$, the following conditions must be satisfied

\begin{align}
 \epsilon_{i,j} S_i S_j = 1. \label{ec:1}
\end{align}

This condition introduces an effective frustration. In such system there will
always be empty sites due to this constraint.

In order to introduce external vibrations and gravity, the square lattice is tilted by $45$ degree. In this way particles diffusing on the lattice will always move up or down since there are no neighboring sites at the same vertical position. The model is used to carry out a Monte Carlo (MC) simulation. In each MC step, we choose a particle and a move type (spin flip or particle hopping) at random. A spin flip is accepted with probability one if there is no violation of Eq. \ref{ec:1}. If a particle attempts a hop, one of the four neighboring sites is chosen with equal probability as the landing site. If the particle attempts a move upward (either up-right or up-left), the landing site is empty and the internal degrees of freedom satisfy Eq.~\ref{ec:1}, we accept the move with probability $P_2$. If the particle attempts to move downward, we accept the move with probability $P_1$ (with $P_1+P_2=1$), provided that Eq.~\ref{ec:1} holds and that the landing site is available. 

In the absence of vibrations, the effect of gravity is simulated by imposing $P_2=0$. When vibrations are switched on, $P_2$ becomes finite. The control parameter is the ratio $x=P_2/ P_1$
which describes the intensity of the vibration.

The MC simulations of the model described above are done on a tilted lattice with periodic boundary conditions along the horizontal axis and rigid walls at bottom and top. The top wall is always high enough to avoid particles to reach it during vibration. After fixing the random bond matrix $\epsilon_{i,j}$, an
initial configuration is prepared by randomly inserting particles of given spin one at a time into the box from its top. Each particle falls down following the described dynamics with $x=0$. After a particle has settled, a new one is released. We introduce a fixed number $N$ of particles. To obtain an initial low density configuration, particle spins are not allowed to flip in this preparation process. The state prepared in this way has a packing fraction of $\phi=0.563 \pm 0.007$. Packing fraction $\phi$ is always measured as the fraction of occupied sites in a rectangular region that span the system in the horizontal direction and whose height is equivalent to the height that the system would have had at $\phi=1$ (i.e., if no voids where present). The region of measurement is vertically centered with the center of mass of the system.

\section{Annealed tapping}
\label{sec:2}

\begin{figure}
\includegraphics[width=0.45\textwidth]{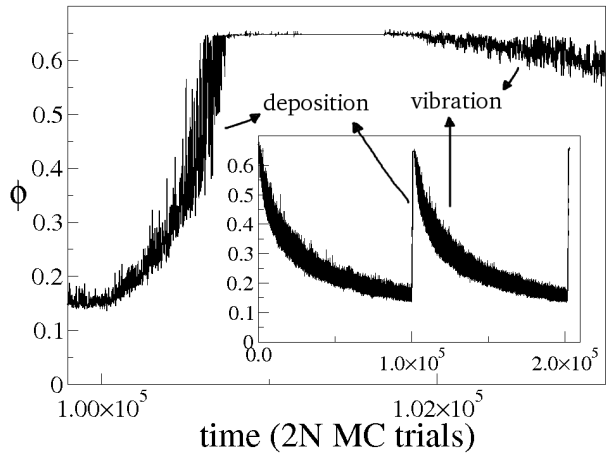}
\caption{Packing fraction $\phi$ as a function of MC time $t$ during the deposition phase in a tap for $x_0=1.0$ and $\tau=10^5$. At $t=10^5$ the vibration is switched off and deposition begins. After stabilization, the vibration is switched back on at $t\approx 1.017\times 10^5$. The inset shows $\phi$ along two full taps. The deposition phases cover a very narrow slot of time as compared with the long duration of the vibration phases in this example.} \label{fig0}
\end{figure}

In the studies carried out by Nicodemi et al. \cite{Nicodemi2}, tapping is simulated through different protocols depending on the shape of the function $x(t)$ used to control vibrations. Here, we use a stepwise function that switches on the vibration to a constant value of $x$ over a period of time $\tau$ and then switches it back off until full deposition and equilibration is reached. Hence, each tap is simulated using

\begin{align}
 x(t)=x_0 [1-\theta(t-\tau)], \label{ec:2}
\end{align}
with $\theta(t)$ the Heaviside step function.

Full deposition and equilibration is defined as the state in which all spins remain in the same site for at least $200$ unit time even though they may flip. A unit time is defined as $2N$ MC trials. In average, each particle has one chance to hop and one chance to flip in this time.

In Fig.~\ref{fig0}, we show the evolution of the packing fraction $\phi$ during a 'strong' tap with $x_0=1.0$ and $\tau=10^5$. Whenever the vibration is on [$x(t)=x_0$], $\phi$ (measured in a central box, as described in Sect.~\ref{sec:1}) fluctuates rapidly but continuously decreases in average. Once vibrations are turned off [$x(t)=0$], $\phi$ rapidly grows as particles fall until full deposition is reached and a constant value of $\phi$ is obtained. We have run some trial simulations in which we wait $1000$ unit time (instead of $200$) after all particle migrations have ceased before deciding that the system is fully stable and have found no differences in the results.

Since we are interested in simulating strong taps, we have investigated different values of the parameters $x_0$ and $\tau$ in order to achieve an effective expansion of the system during vibration. We have found that setting $x_0=1$ and varying $\tau$ allows us to explore a wide range of effective expansions. It is worth mentioning that Nicodemi et al. \cite{Nicodemi1,Nicodemi2} have investigated conditions where the slow relaxation of the system were under scrutiny and for this reason they were mainly focused on values of $x_0$ below $1$. We find that for $x_0=1$ and $\tau < 100$ the system barely expands during the vibration phase. However, for $\tau > 100$ an effective expansion of the system is achieved. We measure the dimensionless expansion $A$ as the ratio between the packing fraction of the system at the time we switch off the vibration ($t=\tau$) and the packing fraction after the bed is fully deposited (see Sect.~\ref{sec:3}).

\begin{figure}
\includegraphics[width=0.45\textwidth]{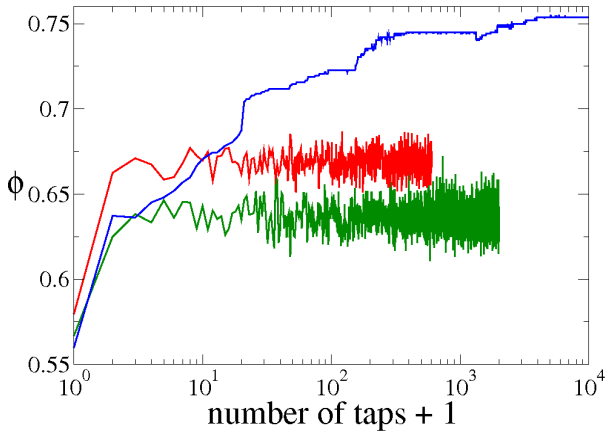}
\caption{Packing fraction $\phi$ as a function of the number of taps applied for
$x_0=1$ and $\tau = 10$ (upper curve), $10^4$ (lower curve), and $1.5\times 10^5$ (middle curve).} \label{fig1}
\end{figure}

In Fig.~\ref{fig1}, we show, as an example, the evolution of the packing fraction $\phi$ during tapping with $x_0=1$ for three values of $\tau$. These correspond to independent experiments with an initial low density configuration prepared as described above. It is clear from the figure that low values of $\tau$ lead to slow relaxation whereas high values of $\tau$ yield a rapid equilibration.

The annealing protocol is carried out as follows. After the initial preparation of the pack, $10^3$ taps with $x_0=1$ are applied at a low tapping intensity, i.e., with a low value of $\tau$. Then, the tapping strength is increased by increasing the value of $\tau$ and a new series of $10^3$ taps are applied to the final configuration reached in the previous series. The process is repeated up to very large values of $\tau$. Then, a decreasing ramp of $\tau$ values is followed down to the initial low tapping intensity.

\section{Results for the high intensity tapping regime}
\label{sec:3}
We use a $30 \times 4000$ squared lattice where we introduce $N=1000$ particles. Initially, particles are ``dropped'' one at a time from above (with random horizontal position and spin) and they move according to the prescribed dynamics with $x=0$ but spin flips are not allowed. Whenever a particle is unable to migrate to a neighboring site after $16$ MC trials the particle is frozen in its site and a new particle is released until all particles are deposited.

The tapping intensity---controlled by $\tau$---is increased and decreased in discrete steps form $\tau=1$ up to $\tau=1.5\times10^5$ and back down to $\tau=1$. At each value of $\tau$, a total of $1000$ taps are applied to the system. The value of $\phi$ considered for each $\tau$ corresponds to the average over the last $500$ taps. The entire annealing is repeated from the initial low density configuration for five different random matrices $\epsilon_{i,j}$. All values of $\phi$ reported correspond to the average over these five independent experiments and error bars display the corresponding standard deviation. We notice that, although the mean value of $\phi$ has a very low dispersion, the ``instantaneous'' $\phi$ in any given run may present larger fluctuations around the mean (see Fig.~\ref{fig1}).

\begin{figure}
\includegraphics[width=0.45\textwidth]{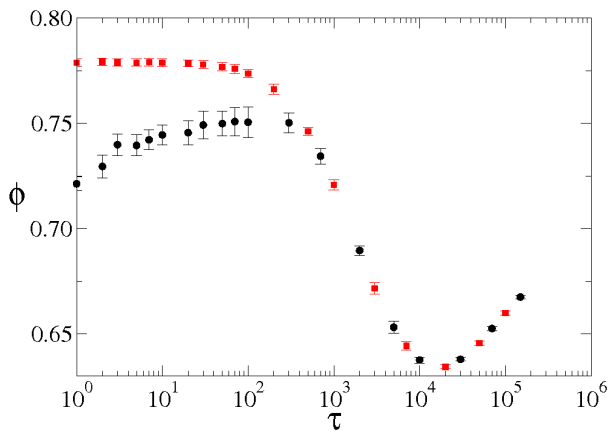}
\caption{Packing fraction $\phi$ as a function of $\tau$ for $x_0=1$ along an annealing protocol. The circles correspond to the ``non-reversible branch'' of initial increasing tapping intensities (increasing $\tau$). The squares correspond to the ``reversible branch'' of decreasing tapping intensities (decreasing $\tau$). The system is tapped $10^3$ times at each value of $\tau$. The last $500$ configurations are averaged to estimate the value of $\phi$ reached. The entire simulation has been repeated five times for different bond random matrices $\epsilon_{i,j}$. The reported data corresponds to the average---and the error bars to the standard deviation---over these five repetitions} \label{fig2}
\end{figure}

In Fig.~\ref{fig2} we show the packing fraction $\phi$ as a function of $\tau$ for the entire annealing protocol. The increasing (irreversible) and decreasing (reversible) branches are indicated with different symbols. We have also increased $\tau$ from the last high density configuration in a second annealing cycle (data not shown). The same values of $\phi$ obtained during the decreasing ramp are reproduced. This confirms that the upper curve shown in Fig.~\ref{fig2} is indeed reversible.

It is clear from Fig.~\ref{fig2} that a minimum in density is present at high values of $\tau$. Beyond this minimum, $\phi$ increases with $\tau$ rather slowly (notice the log scale in the horizontal axis). Due to limitations in CPU time we are unable to explore, at present, this regime beyond this values of $\tau$. However, we can estimate a limiting value of $\phi$ for very strong tapping (see below).

The existence of a minimum in the density--tapping intensity  curve has been reported very recently \cite{Pugnaloni1}. In Ref.~\cite{Pugnaloni1}, the effect is explained in terms of the formation of arches. At low tapping intensities, the introduction of free volume is very limited (small expansion). An increase in effective expansion promotes the formation of larger arches (and hence larger voids) since particles need some space in order to arrange in these cooperative structures. However, if expansion during tapping is too large, falling particles have less chances to meet each other at the time of reaching their stable positions in order to cooperate and form arches.

In the present model arches are not clearly defined. A definition of which particles support a given particle is required to identify arches \cite{Pugnaloni3,Pugnaloni2}. However, the frustration introduced by the bond matrix $\epsilon_{i,j}$ induces the particles to interlock in loops that leave voids in the structure. These loops are the analogous to arches in more realistic models.

\begin{figure}
\includegraphics[width=0.45\textwidth]{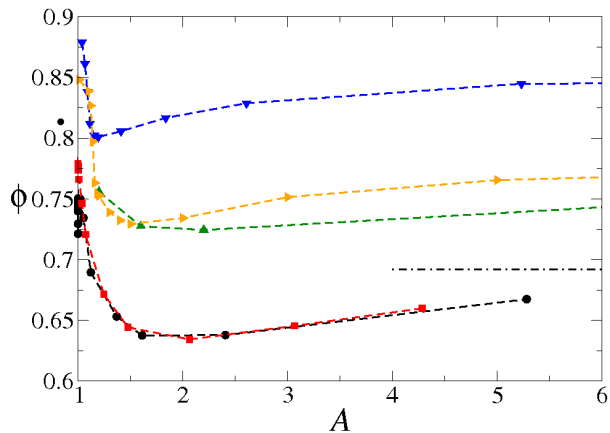}
\caption{Packing fraction $\phi$ as a function of the effective dimensionless expansion $A$ induced by tapping. For each value of $\tau$ the corresponding expansion is calculated as the ratio between the density at the time when the vibration is switched off and the density after full deposition. Symbols as in Fig.~\ref{fig2}. We show results taken from Ref.~\cite{Pugnaloni1} for a molecular dynamics of dissipative soft disks with static friction (down triangles), a pseudo dynamic model of inelastic disks (right triangles) and an off lattice MC of hard disks (up triangles). The horizontal dash-dotted line corresponds to the limit of very strong taps ($\tau \rightarrow \infty$) in the lattice model (see text).} \label{fig3}
\end{figure}

In order to compare our results with previous models, we plot in Fig.~\ref{fig3} the packing fraction as a function of the effective dimensionless expansion $A$ induced by the corresponding value of $\tau$. Results from Ref.~\cite{Pugnaloni1} for twodimensional models are also reproduced. We observe that all models present a minimum even though the position varies in the range $1.10<A<2.10$. The position of the minimum found in 3D models seems to be in the range $3.0<A<5.0$ \cite{Pugnaloni1}. The frustrated lattice gas model yields rather low packing fractions when compared with more realistic models. However, it is clear that the main qualitative features of the reversible branch of the annealed tapping curve are very well captured by the lattice model.

Since the increase in $\phi$ at large values of $\tau$ seems to be due to the particles depositing without much chances to interact as they reach the surface of the growing stable pile, we can simulate this in a extreme condition. Instead of expanding the system using very long $\tau$ values, we now fill the box dropping particles one at time so that they reach the pile individually. This is a process similar to the one used in generating the very first initial condition. However, we now allow the particles to flip their spin as they fall. Moreover, the already deposited particles are not freezed but allowed to diffuse and flip while the pile is being built up. We expect this filling process to mimic the limiting case of an expansion obtained with $\tau \rightarrow \infty$. The packing fraction so obtained is shown in Fig.~\ref{fig3}. We argue that this should be the limiting value to which $\phi$ would approach if we further increased $\tau$.

\section{Conclusions}
\label{sec:4}
We have shown that a well known frustrated lattice gas model of granular compaction---initially designed to study slow dynamics under gentle tapping---is able to reproduce the qualitative features of more complex models in which strong tapping conditions are simulated. In particular, we observe the existence of a minimum in the reversible branch of the annealed tapping. It is somewhat surprising that such simple model can display nonmonotonic behavior. This makes the model a suitable working platform to study this newly discovered feature of granular compaction. It would be of special interest to investigate if the steady states of equal mean packing fraction produced at different tapping intensities (at both sides of the minimum) can be distinguished through the size of the volume fluctuations, an order parameter, or the distribution of characteristic structures such as voids or loops. This model presents the advantage that is of simple implementation and that simulation are much less CPU time demanding than more realistic models which show the same general features.

In Ref.~\cite{Pugnaloni1}, the existence of a minimum in the reversible branch of granular compaction is explained in terms of arching. Since this lattice model does not consider proper contact dynamics nor realistic volume exclusion, it is clear that the presence of the minimum is due to an underlying phenomenon that controls both, arching and packing fraction.

At low tapping intensities, the high packing fractions achieved are due to repeated tapping that allows the system to relax in search for progressively more compact (lower potential energy) structures. This is achieved through the local rearrangements promoted from tap to tap. However, at very high tapping intensities, the system achieves the steady state in a single tap (see Fig.~\ref{fig1}) since the full pack is arranged \emph{de novo} after a strong tap in a global relaxation event. In this last case, in order to achieve higher packing fractions, the particles need to follow a deposition process that allows them enough time to search for the lowest potential energy configuration in a single tap. This is best promoted if the bed is expanded significantly during the tap so that the longer deposition times required and the larger free volume introduced give greater chances for the particles to find lower positions. In summary, the existence of the minimum in the reversible branch of granular compaction seems to be ultimately related to a competition between the global and local relaxation promoted at different scales of tapping strength \cite{Mehta1}.

\begin{acknowledgements}
LAP acknowledge financial support from ANPCyT through project PICT-2007-00882 (Argentina).
\end{acknowledgements}


\end{document}